\documentclass[a4paper]{jpconf}
\usepackage{graphicx}
\usepackage{amssymb}
\usepackage{stmaryrd}
\usepackage{amsmath}
\usepackage{amsfonts}
\usepackage{mathrsfs}

\usepackage{amsmath,amssymb,amsfonts}

\def\be{\begin{equation}}
\def\ee{\end{equation}}
\def\ba{\begin{eqnarray}}
\def\ea{\end{eqnarray}}
\def\nn{\nonumber}

\newcommand{\R}{\mathcal {R}} 
\newcommand{\ints}{{\int_\Sigma}} 

\newcommand{\grav}{\mathrm{gr}} 
\newcommand{\sca}{\mathrm{sc}} 
\newcommand{\kin}{\mathrm{kin}} 
\newcommand{\hil}{\mathcal{H}} 

\begin{document}
\title{Loop Quantum Brans-Dicke Theory}

\author{Xiangdong Zhang and Yongge Ma}

\address{Department of Physics, Beijing Normal University,
Beijing 100875, China}

\ead{zhangxiangdong@mail.bnu.edu.cn; mayg@bnu.edu.cn}

\begin{abstract}
The loop quantization of Brans-Dicke theory (with coupling parameter
$\omega\neq-\frac{3}{2}$) is studied. In the geometry-dynamical
formalism, the canonical structure and constraint algebra of this
theory are similar to those of general relativity coupled with a
scalar field. The connection dynamical formalism of the Brans-Dicke
theory with real $su(2)$-connections as configuration variables is
obtained by canonical transformations. The quantum kinematical
Hilbert space of Brans-Dicke theory is constituted as of that loop
quantum gravity coupled with a polymer-like scalar field. The
Hamiltonian constraint is promoted as a well defined operator to
represent quantum dynamics. This formalism enable us to extend the
scheme of non-perturbative loop quantum gravity to the Brans-Dicke
theory.
\end{abstract}

\section{Introduction}
In the past 25 years, loop quantum gravity(LQG), a background
independent approach to quantize general relativity (GR), has been
widely investigated \cite{Ro04,Th07,As04,Ma07}. Recently, this
non-perturbatively loop quantization procedure has been generalized
to the metric $f(\R)$ theories\cite{Zh11,Zh11b}. In fact, modified
gravity theories have recently received increased attention in
issues related to "dark Universe" and non-trivial tests on gravity
beyond GR. Besides $f(\R)$ theories, a well-known competing
relativistic theory of gravity was proposed by Brans and Dicke in
1961 \cite{BD}, which is apparently compatible with Mach's
principle. To represent a varying "gravitational constant", a scalar
field is non-minimally coupled to the metric in Brans-Dicke
theories(BDT). On the other hand, since 1998, a series of
independent observations implied that our universe is currently
undergoing a period of accelerated expansion\cite{Fr08}. These
results have caused the "dark energy" problem in the framework of
GR. It is reasonable to consider the possibility that GR is not a
valid theory of gravity on a galactic or cosmological scale. The
scalar field in BDT of gravity is then expected to account for "dark
energy". Furthermore, a large part of the non-trivial tests on
gravity theory is related to Einstein's equivalence principle (EEP)
\cite{will}. There exist many local experiments in solar-system
supporting EEP, which implies the metric theories of gravity.
Actually, BDT are a class of representative metric theories, which
have been received most attention. Thus it is interesting to see
whether this class of metric theories of gravity could be quantized
nonperturbatively. Note that the metric $f(\R)$ theories are
equivalent to the special kind of BDT with the coupling parameter
$\omega=0$ and some non-vanishing potential of the scalar
field\cite{So}. In this work, for simplicity consideration, we only
consider BDT with coupling parameter $\omega\neq-\frac{3}{2}$. The
connection formalism of BDT is derived from its geometrical
dynamics. Based on the resulted connection dynamical formalism, we
then quantize the BDT by extending the nonperturbative quantization
procedure of LQG in the way similar to loop quantum $f(\R)$ gravity.
Throughout the paper, we use Greek alphabet for spacetime indices,
Latin alphabet a,b,c,..., for spatial indices, and i,j,k,..., for
internal indices.

\section{Classical and Quantum Aspects of Brans-Dicke Theories }
The original action of Brans-Dicke theories reads \ba
S(g)=\frac12\int_\Sigma
d^4x\sqrt{-g}[\phi\R-\frac{\omega}{\phi}(\partial_\mu\phi)\partial^\mu\phi]\label{action}
\ea where we set $8\pi G=1$, $\R$ denotes the scalar curvature of
spacetime metric $g_{\mu\nu}$, The Hamiltonian analysis of BDT can
be found in Refs.\cite{Zh11c,olmo}. By doing 3+1 decomposition of
the spacetime, the four-dimensional scalar curvature can be
expressed as \ba\mathcal
{R}=K_{ab}K^{ab}-K^2+R+\frac{2}{\sqrt{-g}}\partial_\mu(\sqrt{-g}n^\mu
K)-\frac{2}{N\sqrt{h}}\partial_a (\sqrt{h}h^{ab}\partial_bN),\ea
where $K_{ab}$ is the extrinsic curvature of a spatial hypersurface
$\Sigma$, $K\equiv K_{ab}h^{ab}$, $R$ denotes the scalar curvature
of the 3-metric $h_{ab}$ induced on $\Sigma$, $n^\mu$ is the unit
normal of $\Sigma$ and $N$ is the lapse function. By Legendre
transformation, the momenta conjugate to the dynamical variables
$h_{ab}$ and $\phi$ are defined respectively as \ba
p^{ab}&=&\frac{\partial\mathcal
{L}}{\partial\dot{h}_{ab}}=\frac{\sqrt{h}}{2}[\phi(K^{ab}-Kh^{ab})-\frac{h^{ab}}{N}(\dot{\phi}-N^c\partial_c\phi)], \label{04}\\
\pi&=&\frac{\partial\mathcal
{L}}{\partial\dot{\phi}}=-\sqrt{h}(K-\frac{\omega}{N\phi}(\dot{\phi}-N^c\partial_c\phi)),\label{pi}
\ea  where $N^c$ is the shift vector. The resulted Hamiltonian of
BDT can be derived as a liner combination of constraints as
$H_{total}=\int_\Sigma d^3x(N^aV_a+NH),$ where the smeared
diffeomorphism and Hamiltonian constraints read respectively \ba
V(\overrightarrow{N})&=&\int_\Sigma d^3xN^aV_a =\int_\Sigma
d^3xN^a\left(-2D^b(p_{ab})+\pi\partial_a\phi\right),\label{dc}\\
H(N) &=&\int_\Sigma
d^3xN\left[\frac2{\sqrt{h}}\left(\frac{p_{ab}p^{ab}-\frac12p^2}{\phi}+\frac{(p-\phi\pi)^2}{2\phi(3+2\omega)}\right)
+\frac12\sqrt{h}(-\phi R+\frac{\omega}{\phi}(D_a\phi)
D^a\phi+2D_aD^a\phi)\right].\label{hc}\nn\\\ea Here the condition
$\omega\neq-\frac32$ was assumed. Lengthy but straightforward
calculations show that the constraints comprise a first-class system
similar to GR. Since the geometric canonical variables
$(h_{ab},p^{ab})$ of BDT are as same as those of $f(\R)$ theories
\cite{Zh11b}, we can use the same canonical transformations of
$f(\R)$ theories to obtain the connection dynamical formalism of
BDT. Let \ba\tilde{K}^{ab}=\phi
K^{ab}+\frac{h^{ab}}{2N}(\dot{\phi}-N^c\partial_c\phi).\ea The new
geometric variables are $E^a_i=\sqrt{h}e^a_i$ and $
A^i_a=\Gamma^i_a+\gamma\tilde{K}^i_a,$ where $e^a_i$ is the triad
such that $h_{ab}e^a_ie^b_j=\delta_{ij}$,
$\tilde{K}^a_i\equiv\tilde{K}^{ab}e_b^i$, $\Gamma^i_a$ is the spin
connection determined by $E^a_i$, and $\gamma$ is a nonzero real
number. It is clear that our new variable $A^i_a$ coincides with the
Ashtekar-Barbero connection \cite{As86,Ba} when $\phi=1$. The only
non-zero Poisson brackets among the new variables reads
$\{A^j_a(x),E_k^b(y)\}=\gamma\delta^b_a\delta^j_k\delta(x,y).$ Now,
the phase space of BDT consists of conjugate pairs $(A_a^i,E^b_j)$
and $(\phi,\pi)$, with the additional Gaussian constraint $\mathcal
{G}_i=\mathscr{D}_aE^a_i\equiv\partial_aE^a_i+\epsilon_{ijk}A^j_aE^{ak},
$ which justifies $A^i_a$ as an $su(2)$-connection. The original
vector and Hamiltonian constraints can be respectively written up to
Gaussian constraint as \ba V_a &=&\frac1\gamma
F^i_{ab}E^b_i+\pi\partial_a\phi, \\
H&=&\frac{\phi}{2}\left[F^j_{ab}-(\gamma^2+\frac{1}{\phi^2})\varepsilon_{jmn}\tilde{K}^m_a\tilde{K}^n_b\right]
\frac{\varepsilon_{jkl}
E^a_kE^b_l}{\sqrt{h}}\nn\\
&+&\frac1{3+2\omega}\left(\frac{(\tilde{K}^i_aE^a_i)^2}{\phi\sqrt{h}}+
2\frac{(\tilde{K}^i_aE^a_i)\pi}{\sqrt{h}}+\frac{\pi^2\phi}{\sqrt{h}}\right)
+\frac{\omega}{2\phi}\sqrt{h}(D_a\phi)
D^a\phi+\sqrt{h}D_aD^a\phi,\label{hamilton} \ea where
$F^i_{ab}\equiv2\partial_{[a}A^i_{b]}+\epsilon^i_{kl}A_a^kA_b^l$ is
the curvature of $A_a^i$. All the constraints are of first class.
The total Hamiltonian can be expressed as a linear combination
$H_{total}=\ints\Lambda^i\mathcal {G}_i+N^aV_a+NH.$

Based on the connection dynamical formalism, the nonperturbative
loop quantization procedure can be straightforwardly extended to the
BDT. The kinematical structure of BDT is as same as that of $f(\R)$
theories \cite{Zh11,Zh11b}. The kinematical Hilbert space of the
system is a direct product of the Hilbert space of geometry and that
of scalar field, $\hil_\kin:=\hil^\grav_\kin\otimes \hil^\sca_\kin$,
with the orthonormal spin-scalar-network basis
$T_{\alpha,X}(A,\phi)\equiv T_{\alpha}(A)\otimes T_{X}(\phi)$ over
some graph $\alpha\cup X\subset\Sigma$. Here $\alpha$ and $X$
consist of finite number of curves and points respectively in
$\Sigma$. The basic operators are the quantum analogue of holonomies
$h_e(A)=\mathcal {P}\exp\int_eA_a$ of connections, densitized triads
smeared over 2-surfaces $E(S,f):=\int_S\epsilon_{abc}E^a_if^i$,
point holonomis $U_\lambda=\exp(i\lambda\phi(x))$\cite{As03}, and
scalar momenta smeared on 3-dimensional regions $\pi(R):=\int_R
d^3x\pi(x)$. Note that the whole construction is background
independent, and the spatial geometric operators of LQG, such as the
area \cite{Ro95}, the volume \cite{As97} and the length operators
\cite{Th98,Ma10} are still valid here. As in LQG, it is
straightforward to promote the Gaussian constraint $\mathcal
{G}(\Lambda)$ to a well-defined operator\cite{Th07,Ma07}. It's
kernel is the internal gauge invariant Hilbert space $\mathcal
{H}_G$ with gauge invariant spin-scalar-network basis. Since the
diffeomorphisms of $\Sigma$ act covariantly on the cylindrical
functions in $\mathcal {H}_G$, the so-called group averaging
technique can be employed to solve the diffeomorphism
constraint\cite{As04,Ma07}. Thus we can also obtain the desired
diffeomorphism and gauge invariant Hilbert space $\mathcal
{H}_{Diff}$ for the BDT.

Now, we come to implement the Hamiltonian constraint
(\ref{hamilton}) at quantum level. In order to compare the
Hamiltonian constraint of BDT with that of $f(\R)$ theories in
connection formalism, we write Eq. (\ref{hamilton}) as
$H(N)=\sum^7_{i=1}H_i$. It is easy to see that the terms
$H_1,H_2,H_7$ just keep the same form as those in $f(\R)$ theories,
the $H_3,H_4,H_5$ terms are also similar to the corresponding terms
in $f(\R)$ theories. Here differences are only reflected by the
coefficients. Now we come to the completely new term,
$H_6=\int_\Sigma d^3xN\frac{\omega}{2\phi}\sqrt{h}(D_a\phi) D^a\phi
$. We can introduce well-defined operators $\phi,\phi^{-1}$ as in
Ref. \cite{Zh11b}. By the same regularization techniques as in
Refs.\cite{Zh11b,Ma06}, we triangulate $\Sigma$ in adaptation to
some graph $\alpha$ underling a cylindrical function in $\hil_\kin$
and reexpress connections by holonomies. The corresponding regulated
operator $\hat{H}^\varepsilon_6$ can acts on a basis vector
$T_{\alpha,X}$ over some graph $\alpha\cup X$. It is easy to see
that the action of $\hat{H}^\varepsilon_6$ on $ T_{\alpha,X}$ is
graph changing. It adds a finite number of vertices at
$t(s_I(v))=\varepsilon$ for edges $e_I(t)$ starting from each
high-valent vertex of $\alpha$. As a result, the family of operators
$\hat{H}^\varepsilon_6(N)$ fails to be weakly convergent when
$\varepsilon\rightarrow 0$. However, due to the diffeomorphism
covariant properties of the triangulation, the limit operator can be
well defined via the so-called uniform Rovelli-Smolin topology
induced by diffeomorphism-invariant states $\Phi_{Diff}$. It is
obviously that the limit is independent of $\varepsilon$. Hence the
regulators can be removed. We then have \ba \hat{H}_6\cdot
T_{\alpha,X} &=&\sum_{v\in
V(\alpha)}\frac{2^{17}N(v)\omega }{3^6\gamma^4(i\lambda_0)^2(i\hbar)^4E^2(v)}\hat{\phi}^{-1}(v)\nn\\
&\times&\sum_{v(\Delta)=v(\Delta')=v}\epsilon(s_L s_M
s_N)\epsilon^{LMN}\hat{U}^{-1}_{\lambda_0}(\phi(s_{s_L(\Delta)}))
[\hat{U}_{\lambda_0}(\phi(t_{s_L(\Delta)}))-\hat{U}_{\lambda_0}(\phi(s_{s_L(\Delta)}))]\nn\\
&\times&\Tr(\tau_i\hat{h}_{s_M(\Delta)}[\hat{h}^{-1}_{s_M(\Delta)},(\hat{V}_v)^{3/4}]
\hat{h}_{s_N(\Delta)}[\hat{h}^{-1}_{s_N(\Delta)},(\hat{V}_v)^{3/4}]) \nn\\
&\times&\epsilon(s_I s_J
s_K)\epsilon^{IJK}\hat{U}^{-1}_{\lambda_0}(\phi(s_{s_I(\Delta')}))
[\hat{U}_{\lambda_0}(\phi(t_{s_I(\Delta')}))-\hat{U}_{\lambda_0}(\phi(s_{s_I(\Delta')}))]\nn\\
&\times&\Tr(\tau_i\hat{h}_{s_J(\Delta')})[\hat{h}^{-1}_{s_J(\Delta')},(\hat{V}_{v})^{3/4}]
\hat{h}_{s_K(\Delta')}[\hat{h}^{-1}_{s_K(\Delta')},(\hat{V}_{v})^{3/4}])\cdot
T_{\alpha,X} . \ea Thus the total Hamiltonian constraint operator
$\hat{H}(N)=\sum^7_{i=1}\hat{H}_i$ is well defined in $\hil_G$.
Furthermore, master constraint programme can be introduced for BDT
to avoid possible quantum anomaly and find the physical Hilbert
space\cite{Zh11c}.

\section{Conclusions}
With the key observation that LQG is based on its $su(2)$-connection
dynamical formalism which can be derived via canonical
transformations from the geometric dynamics, the $su(2)$-connection
dynamics of BDT is obtained. Thus LQG has been successfully extended
to the BDT by coupling to a polymer-like scalar field. The quantum
kinematical structure of BDT is as same as that of loop quantum
$f(\R)$ theories. Hence the important physical result that both the
area and the volume are discrete remains valid for quantum BDT.
While the dynamics of BDT is more general than that of $f(\R)$
theories, the Hamiltonian constraint can still be promoted to a
well-defined operator in $\hil_G$. Hence the classical BDT can be
non-perturbatively quantized. Therefore, besides GR and $f(\R)$
theories, LQG method is also valid for the BDT of gravity.

\ack We thank the organizers of Loops 11 conference for the
financial support to our attendance. This work is supported by NSFC
(Grant No.10975017) and the Fundamental Research Funds for the
Central Universities.

\end{document}